\documentclass[final,5p,times,twocolumn]{elsarticle}

\usepackage{bm}
\usepackage{amsmath}
\usepackage{fleqn}
\begin{document}

\begin{frontmatter}



\title{On the Chudakov effect for the most probable value of high-energy electron-positron pair ionization loss in thin targets}


\author{S.V. Trofymenko}
\ead{strofymenko@gmail.com}
\address{National Science Center `Kharkiv Institute of Physics and Technology', 1 Akademichna st., 61108 Kharkiv, Ukraine}
\address{Karazin Kharkiv National University, 4 Svobody sq., 61022 Kharkiv, Ukraine}




\begin{abstract}
The ionization loss of a high-energy electron-positron pair in thin targets is considered. The analogue of the Landau distribution function is derived for this loss under the condition when the Chudakov effect of the pair ionization loss suppression is manifested. Expression for the most probable value of the pair ionization loss $E_{MP}$ is obtained. It is shown that the magnitude of Chudakov effect for $E_{MP}$ can be noticeably different from the magnitude of this effect for the restricted mean value of the pair ionization loss. 



\end{abstract}


\end{frontmatter}


\section{Introduction}
\label{Introduction}

The value of ionization energy loss of a fast particle traversing thin target is stochastic. It is distributed according to the law firstly derived by Landau \cite{Landau1944}. Such a distribution, known as straggling function, is asymmetric with respect to its single maximum corresponding to the most probable value of the ionization loss $E_{MP}$. This value is one of the key parameters of the straggling function. It is smaller than the average value $E_{AV}$ of the particle ionization loss in the target, defined by the Bethe-Bloch formula. $E_{MP}$ depends a bit differently on the target thickness $x$ than $E_{AV}$. Namely, $E_{MP}\propto x(\ln x+a)$, where $a$ is some constant, and grows faster than linearly with the increase of the target thickness, while, naturally, $E_{AV}\propto x$. Straggling functions of various particles in amorphous targets or disoriented crystals have been extensively studied both theoretically \cite{BlunckLeisegang1950,Vavilov1957,Shulek1967,Talman1979,Bichsel1988} and experimentally \cite{AllisonCobb1980,Bak1987}. A series of studies has been performed for the case of particle penetration through oriented crystals \cite{Esbensen1978,PapeMoller2001,TrofymenkoKyryllin2020} where the straggling functions have a more complicated structure than in amorphous targets. 

In the above works it was the ionization loss by single particles that was investigated, i. e. the loss under the condition when the particles traverse the target one at a time. As far as the quantity $E_{AV}$ is concerned, a series of works has been devoted to the study of the influence of interference effects on its value when several particles penetrate the target simultaneously. Probably, the most well-known effect of this kind is the Chudakov effect (or King-Perkins-Chudakov effect) \cite{Chudakov,Perkins}. It is the suppression of ionization energy loss of a high-energy electron-positron pair in the vicinity of its creation point in substance compared to the sum of independent electron and positron ionization energy losses. It occurs due to mutual screening (destructive interference) of the electron's and positron's proper Coulomb fields. Due to polarization of the substance the typical transversal scale of these fields is $1/\omega_p$, where $\omega_p$ is the plasma frequency\footnote{We use the system of units where the speed of light $c$ and the Planck constant $\hbar$ are equal to unity.}. The scale of the pair divergence angle is $m/\mathcal{E}\ll1$, where $m$ is the electron mass and $\mathcal{E}$ is the pair energy. Thus, the electron's and positron's proper fields can interfere within the distance $l\sim\mathcal{E}/m\omega_p$ from the creation point, which is the region where the Chudakov effect is manifested. The analogous effect has been also considered for a proton cluster, bound by a common electron, which experiences a ``Coulomb explosion" after its entrance into a target \cite{Brandt,ShulgaSyshchenko1992}. In this case, however, one deals with constructive interference of the protons' fields, which results in the enhancement of the disintegrated cluster energy loss compared to the sum of independent energy losses of separate protons. 

The Chudakov effect has been studied experimentally with the use of cosmic ray photons for pair production in emulsions \cite{Perkins,WolterMiesowicz1956,Iwadare1958}, which showed a fairly good agreement with the theories of this effect for the mean ionization loss of the pair \cite{Chudakov,BerestetskiiGeshkenbain,Yekutieli,MitoEzawa,Burkhardt}. In \cite{Vircus,Thomsen2011} there was reported the first proof-of principle experiment of this kind, where the photons were generated by high-energy electrons in an accelerator. It was noted that such an approach allows for much more precise determination of the pair energy and achievement of better statistics. In this experiment the pairs were created by multi-GeV photons impinging on thin upstream golden foils, while the pair ionization loss was registered in a thin downstream silicon detector with the active layer thickness of 16 $\mu$m situated on some distance from the foils. In this case it was the most probable value of the pair ionization loss $E_{MP}$ that was measured. It was defined from the straggling functions of the pairs inside the detector. The obtained results showed some discrepancy with the theories \cite{Chudakov,BerestetskiiGeshkenbain,Yekutieli,MitoEzawa,Burkhardt} of the mean ionization loss of the pair. Namely, a more significant suppression of the pair ionization loss, than the one expected from these theories, was registered.  

In the present Letter we obtain the straggling function of electron-positron pairs in thin targets taking into account the discussed above interference effect in the pair ionization loss and calculate the corresponding value of the most probable ionization loss $E_{MP}$. The consideration is performed both with the use of approach of \cite{Landau1944}, which allows obtaining analytical expression for $E_{MP}$, and with the use of a more rigorous approach, which provides more accurate results for straggling functions in very thin targets. It is shown that in sufficiently thin targets the magnitude of Chudakov effect is noticeably different for the quantities $E_{MP}$ and $E_{AV}$, and the value $E_{MP}$ is stronger suppressed due to this effect than $E_{AV}$. Therefore, it is essential to use the calculated value of $E_{MP}$, rather than $E_{AV}$, for comparison with experimental results on the most probable values of the pair ionization loss in thin detectors. 


\section{Targets of moderate thickness}
\label{sec2}

In order to obtain the straggling function for $e^+e^-$ pairs in a target of thickness $x$ we follow the approach developed in the seminal paper \cite{Landau1944} for single particles. Let $w(\varepsilon)d\varepsilon$ be the probability for the pair to lose energy in the interval $(\varepsilon,\varepsilon+d\varepsilon)$ within the unit path. In this case the kinetic equation for the straggling function of the pair is completely analogous to the one describing such a function for single particles:
\begin{equation}
\frac{df(x,\Delta)}{dx}=\int\limits_{0}^{\varepsilon_{\textrm{max}}}d\varepsilon w(\varepsilon)[f(x,\Delta-\varepsilon)-f(x,\Delta)],
\label{kin_eq}
\end{equation} 
where $\varepsilon_{\textrm{max}}$ is the maximum energy which the pair can transfer to an atomic electron in a single collision, $\Delta$ is the energy lost by the pair in the target. The straggling function $f(x,\Delta)$ is different from zero only for positive $\Delta$. The solution of this equation with the initial condition $f(0,\Delta)=\delta(\Delta)$ reads
\begin{equation}
f(x,\Delta)=\frac{1}{2\pi i}\int\limits_{-i\infty+a}^{+i\infty+a}dp\exp\bigg\{p\Delta-x\int\limits_{0}^{\varepsilon_{\textrm{max}}}d\varepsilon w(\varepsilon)(1-e^{-p\varepsilon})\bigg\},
\label{f_general}
\end{equation}
where $a$ can be put equal zero and integration with respect to $p$ performed along the imaginary axis. The integral with respect to $\varepsilon$ can be split into two parts and presented as
\begin{equation}
\int\limits_{0}^{\varepsilon_1}d\varepsilon w(\varepsilon)(1-e^{-p\varepsilon})+\int\limits_{\varepsilon_1}^{\varepsilon_{\textrm{max}}}d\varepsilon w(\varepsilon)(1-e^{-p\varepsilon}).
\label{e_integral}
\end{equation}  
The quantity $\varepsilon_1$ should satisfy the condition $\varepsilon_k\ll\varepsilon_1\ll\varepsilon_{\textrm{max}}$, where $\varepsilon_k$ are the binding energies of the atomic electrons. This allows expanding the exponent in the first integral up to the term linear in small  parameter $p\varepsilon$. This integral becomes equal to the mean value of the pair ionization loss per unit path due to collisions with energy transfer not exceeding $\varepsilon_1$, multiplied by $p$. For this value we apply the result from \cite{BerestetskiiGeshkenbain} and obtain  
\begin{equation}
\begin{aligned}
&\int\limits_{0}^{\varepsilon_1}d\varepsilon w(\varepsilon)(1-e^{-p\varepsilon})\approx p\int\limits_{0}^{\varepsilon_1}d\varepsilon\varepsilon w(\varepsilon)\\
&=\frac{p\eta}{x}\bigg\{\ln\frac{2m\varepsilon_1}{\omega_p^2}-2K_0(\omega_p s)\bigg\},
\label{e_integral_1}
\end{aligned}
\end{equation}   
where $\eta=4\pi ne^4x/m$ with $n$ being the number of atomic electrons per unit volume. In terms of conventional denominations applied in the papers on straggling functions $\eta=2\xi$.  Numerical estimations will be further made for silicon targets. In this case $\eta/x\approx0.0356~\textrm{keV}/\mu$m. Here also $K_0$ is the Macdonald function and $s$ is the distance between the electron and the positron in the direction orthogonal to the average velocity of the pair. Following most of the papers on the theory of Chudakov effect, we consider the simplest case when the electron and positron energies are the same. 

In order to calculate the second integral in (\ref{e_integral}) one needs the expression for $w(\varepsilon)$ in the range of $\varepsilon$ where atomic electrons can be considered as free (since $\varepsilon_1\gg \varepsilon_k$). It can be expressed in terms of differential cross section of the energy transfer from the pair to the electron as $w(\varepsilon)=nd\sigma/d\varepsilon$. Let us first calculate the cross section $d\sigma/dq$ differential in the transverse momentum transfer. Being an invariant quantity, it can be calculated in the pair rest frame. Let us neglect the deflection of the high-energy particles constituting the pair from the straight lines during the pair motion in the medium. In this case in the discussed frame of reference one deals with a free atomic electron scattering on the static potential of the pair in a polarized medium\footnote{The electron-positron separation $s$, which is also an invariant, is supposed to remain almost constant during the interaction time.}. This potential can be found by the Lorentz transformation of the corresponding scalar and vector potentials in the laboratory frame. For a single particle of the pair, which moves along the $z$-axis, they read \cite{Akhiezer}: 
\begin{equation}
\varphi_L=\frac{e}{\sqrt{\rho^2\gamma^{-2}+(z-vt)^2}}e^{-\omega_p\sqrt{\rho^2+\gamma^2(z-vt)^2}},~~{\bf A}_L={\bf v}\varphi_L,
\label{potentials}
\end{equation}   
where $v=|{\bf v}|$ is the particle velocity, $\rho$ is the distance from the $z$-axis to the observation point, $\gamma$ is the particle Lorentz-factor. Let us neglect the divergence of the pair and consider the electron and positron as moving in parallel directions. In this case the pair rest frame coincides with the rest frame of each particle. The transformation of (\ref{potentials}) to the pair rest frame gives
\begin{equation}
\varphi({\bf r})=\frac{e}{r}e^{-\omega_p r},~~{\bf A}=0,
\end{equation}   
with $r$ being the distance from the particle. The transition amplitude between the initial and final states of the incident atomic electron reads
\begin{equation}
S_{fi}=-ie\int d^4x\overline\psi_f(x)\varphi_p(x)\gamma_0\psi_i(x),
\label{S_fi}
\end{equation}   
where $\gamma_0$ is the gamma matrix, $\psi_{i,f}$ are the plane waves of the initial and final electron states, $\varphi_p$ is the scalar potential of the pair with the Fourier component
\begin{equation}
\varphi_p({\bf q})=\frac{4\pi e}{q^2+\omega_p^2}\big(1-e^{-i{\bf qs}}\big).
\label{fi_Fourier_component}
\end{equation} 
The vector ${\bf s}$ is directed from the electron (with the charge $e$) to the positron. On the basis of (\ref{S_fi}) and (\ref{fi_Fourier_component}) one obtains the required cross section:
\begin{equation}
\frac{d\sigma}{dq}=16\pi e^4\frac{q}{\big(q^2+\omega_p^2\big)^2}\big[1-J_0(qs)\big],
\end{equation}  
where $J_0$ is the Bessel function. Here we neglected the term proportional to $q^2/p_i^2$, where $p_i$ is the electron initial momentum. It is applicable due to the condition $\eta\ll\varepsilon_{\textrm{max}}$, which we assume to be valid. The obtained cross section has the same form in the laboratory frame due to invariance of $d\sigma/dq$ itself, as well as of all the quantities which it includes (from the very beginning the value of $\omega_p$ refers to the laboratory frame). Applying $\varepsilon=q^2/2m$ one finally obtains the cross section of the pair energy transfer to an atomic electron at $\varepsilon\gg\varepsilon_k$ and the corresponding probability $w(\varepsilon)$:
\begin{equation}
w(\varepsilon)=\frac{\eta}{x}\frac{1-J_0\big(\sqrt{2m\varepsilon}s\big)}{\big(\varepsilon+\omega_p^2/2m\big)^2}.
\label{w_e}
\end{equation}   
In the integrand of the second integral in (\ref{e_integral}) this expression for $w(\varepsilon)$ can be considerably simplified. Indeed, the quantity $\omega_p^2/2m$ is much smaller than 1 eV and can be neglected compared to $\varepsilon\gg\varepsilon_k$. Moreover, for the mean ionization potential of silicon $I=173$ eV, $\sqrt{2mI}\sim7.5\cdot10^8$ cm$^{-1}$. Since $\varepsilon\gg I$, the argument of $J_0$ is large even if $s$ is on the order of interatomic distance (which is the minimum value of $s$ we will be interested in). Thus, one can neglect $J_0$ in (\ref{w_e}) as well. As a result, the second integral in (\ref{e_integral}) can be calculated in the same way as in \cite{Landau1944}, since the simplified expression for $w(\varepsilon)$ here becomes just twice as large as the corresponding expression used in \cite{Landau1944} for single particles. 

Substituting the obtained result for (\ref{e_integral}) into (\ref{f_general}) and integrating here along the imaginary axis, one can present the distribution function in the form, analogous to the one it has for single particles:
\begin{equation}
f(x,\Delta)=\frac{1}{\pi\eta}\int\limits_0^{+\infty}dye^{-\pi y/2}\cos(y\lambda+y\ln y),
\label{f_Landau}
\end{equation}     
where
\begin{equation}
\begin{aligned}
&\lambda=\frac{\Delta-\eta\ln(\eta/\varepsilon')}{\eta}-1+\varGamma,~~ \varGamma~\textrm{is the Euler's constant},\\
&\ln\varepsilon'=\ln(\omega_p^2/2m)+2K_0(\omega_ps).
\label{Lambda}
\end{aligned}
\end{equation}  
It differs from the result \cite{Landau1944} for single particles by the substitution $\xi\to\eta=2\xi$ and a different expression for $\ln\varepsilon'$.

As a function of $\lambda$, the integral in (\ref{f_Landau}) has a maximum at $\lambda\approx-0.223$. Thence, applying (\ref{Lambda}), one obtains the most probable value of the ionization loss $\Delta$ of $e^+e^-$ pair in a thin target:
\begin{equation}
E_{MP}(s)=\eta\bigg\{\ln\frac{2m\eta}{\omega_p^2}-2K_0(\omega_ps)+0.2\bigg\}.
\label{E_mp}
\end{equation}   
Substituting here $z=s/\vartheta$, where $\vartheta$ is the pair divergence angle, one can express $E_{MP}$ in terms of distance $z$ from the pair creation point. Note that expression (\ref{E_mp}) is valid for a pair of sufficiently high energy ($\mathcal{E}\gtrsim 100$ MeV), for which the full value density effect takes place in the ionization loss. In this case $E_{MP}$ does not depend on the energy of the pair. The result (\ref{E_mp}) differs from the corresponding value $E^{S}_{MP}$ for single particles by the substitution $\xi\to\eta$ and the term with $K_0$. Due to the presence of $\eta$ in the argument of the logarithm, for $s\gg\omega_p^{-1}$ the value (\ref{E_mp}) is larger than $E^{S}_{MP}$ by a factor which exceeds 2. Thus, at large separations between the electron and positron the most probable value of the pair ionization loss slightly exceeds the sum of independent $E^{S}_{MP}$ values for each particle of the pair (if the particles would have traversed the target separately). This makes $E_{MP}$ different from the mean value $E_{AV}$ of the pair ionization loss, which for $s\gg\omega_p^{-1}$ becomes equal to the sum of independent electron and positron mean ionization losses.

The fact that for large $s$ the value of $E_{MP}$ exceeds the doubled value of $E^{S}_{MP}$ can be illustrated by the following consideration. Let the straggling function for a single electron in a target of thickness $x$ be $f(x,\Delta)$. The straggling function $f(2x,\Delta)$ in the target of thickness $2x$ is defined by a convolution \cite{Bichsel1988}
\begin{equation}
f(2x,\Delta)=\int\limits_{0}^{\Delta} f(x,\epsilon)f(x,\Delta-\epsilon)d\epsilon,
\label{Convolution}
\end{equation}
since the particle ionization losses in each layer of thickness $x$, which comprise the target, are independent. As noted, the value of $E^{S}_{MP}$, corresponding to this function, exceeds the doubled value of $E^{S}_{MP}$ corresponding to the function $f(x,\Delta)$. For $s\gg\omega_p^{-1}$ the ionization losses of the electron and positron of the pair are independent. Thus, for such $s$ the ionization loss of the pair in a target of thickness $x$ is analogous to the ionization loss of a single electron\footnote{Presently we do not distinguish between the struggling functions of high-energy electrons and positrons, which is justified at least for $\Delta\sim E^{S}_{MP}$.} in the target of thickness $2x$, and the straggling functions in these cases are the same. Therefore, for $s\gg\omega_p^{-1}$ the most probable value $E_{MP}$ of the pair ionization loss in a target of thickness $x$ should be equal to the corresponding value for a single electron in a target of thickness $2x$. Hence, $E_{MP}$ should exceed the doubled value of the most probable ionization loss $E^{S}_{MP}$ for a single electron in a target of thickness $x$.  

Concerning the mean value $E_{AV}$ of the pair ionization loss, with the logarithmic accuracy it is defined as \cite{BerestetskiiGeshkenbain}
\begin{equation}
E_{AV}(s)=\eta\bigg\{\ln\frac{2m\varepsilon_\textrm{max}}{\omega_p^2}-2K_0(\omega_p s)\bigg\}.
\label{E_av}
\end{equation}
For our estimations we take $\varepsilon_\textrm{max}$ equal to the electron (or positron) energy $\gamma m$. In this case expression (\ref{E_av}) differs from the exact expression for the mean total value of the pair ionization loss by a term on the order of unity in braces, which is small compared to the logarithm at high energy of the pair. Further we will make numerical estimations of $E_{AV}$ for the pair energy of $\mathcal{E}=1$ GeV.

For thin targets it might be more convenient to consider not the mean total value of the ionization loss $E_{AV}$, but the restricted one $E_{AV(R)}$. It is a part of the value $E_{AV}$ due to collisions with energy transfer not exceeding some value $\varepsilon_R$, which is smaller than the maximum energy transfer $\varepsilon_\textrm{max}$ defined by the collision kinematics. It is caused by the fact that fast delta-electrons originating from the collisions with a large energy transfer $\varepsilon>\varepsilon_R$ may leave the target without depositing a considerable amount of their energy inside it. The energy transferred by the incident particle to such electrons will not be registered in this case. The value of $E_{AV(R)}$ is defined by the expression (\ref{E_av}) with the substitution $\varepsilon_\textrm{max}\to\varepsilon_R$. The value of $\varepsilon_R$ is usually chosen to be on the order of the energy at which the electron range coincides with the target thickness. 

In order to investigate the magnitude of Chudakov effect for $E_{MP}$ it is convenient to consider the ratio of this quantity as a function of $s$ to its value $E_{MP}(\infty)$ at $s\gg\omega_p^{-1}$: $\alpha_{MP}=E_{MP}(s)/E_{MP}(\infty)$. The analogous ratios $\alpha_{AV}=E_{AV}(s)/E_{AV}(\infty)$ and $\alpha_{AV(R)}=E_{AV(R)}(s)/E_{AV(R)}(\infty)$ can be introduced for $E_{AV}$ and $E_{AV(R)}$ as well. Numerical estimations on the basis of (\ref{E_mp}) and (\ref{E_av}) show that generally $\alpha_{AV}>\alpha_{AV(R)}>\alpha_{MP}$ and the difference between these quantities increases with the decrease of the target thickness. The numerical results for $\alpha_{MP}$, $\alpha_{AV}$ and $\alpha_{AV(R)}$ will be presented in the next section (see Fig.~\ref{fig2}) where we treat more accurately the case of very thin targets in which the difference in the magnitude of Chudakov effect for $E_{MP}$ and $E_{AV}$ (or $E_{AV(R)}$) is the most noticeable. Though, as will be noted, expression (\ref{E_mp}) nicely works for calculation of $\alpha_{MP}$ (however, not $E_{MP}$ itself) even in sufficiently thin targets.

\section{Very thin targets}
\label{sec3}

The consideration presented in the previous section, based on the method applied in \cite{Landau1944}, does not properly take into account the fact that atomic electrons are not free but exist in bound states in atoms. As shown (see \cite{Bichsel1988} and refs. therein), for single particles in silicon targets such a consideration is valid for moderate target thickness $x\gtrsim1$ mm. In thinner targets this approach results in a shape of the straggling function which is noticeably different from the one obtained experimentally. Though, its maximum position, corresponding to $E_{MP}$, is not considerably shifted from (\ref{E_mp}) in this case. In the present section we apply a more accurate approach for calculation of a $e^+e^-$ pair straggling function, which is valid also for targets of thickness much less than 1 mm, presently called very thin targets.  

Let us obtain the expression for straggling function on the basis of a more detailed consideration of the cross section of the incident $e^+e^-$ pair energy transfer to an atomic electron. Just like for single incident particles, this cross section can be divided into contributions from close and distant collisions: $\sigma=\sigma_c+\sigma_d$. The distant collision part $\sigma_d$ can be calculated with the use of equivalent photon method, following the approach applied in \cite{Sorensen,Chechin,TrofymenkoPRA2020} for consideration of $K$-shell ionization. In this method the value of $\sigma_d$ per a single atomic electron can be presented as 
\begin{equation}
\sigma_d=\frac{1}{Z}\sum_k\int\limits\frac{dN_k}{d\omega}\sigma^k_{ph}(\omega)d\omega,
\label{sigma_d}
\end{equation} 
where $Z$ is the atomic number of the substance, $\sigma^k_{ph}(\omega)$ is the photoionization cross section of the $k$-th atomic shell, $dN_k/d\omega$ is the spectral density of the equivalent photon number for the electromagnetic field of the pair\footnote{Its value depends on the atomic shell number [see formula (\ref{dNdw})].}. It can be found as
\begin{equation}
\frac{dN_k}{d\omega}=(4\pi^2\omega)^{-1}\int|{\bf E}_\omega|^2d^2\rho,
\label{dNdw_int}
\end{equation}  
where ${\bf E}_\omega$ is the Fourier component of electric field of the pair. Let us again neglect the non-parallelism of the electron and positron trajectories and direct the $z$-axis along the velocities of the particles. In this case, provided the energy of the pair is sufficiently high, ${\bf E}_\omega$ can be presented in the form of the following Fourier expansion \cite{TrofymenkoPLA2013}: 
\begin{equation}
{\bf E}_\omega=-\frac{ie}{\pi}e^{i\frac{\omega}{v}z}\int d^2q\frac{\bf q}{q^2+\Omega^2}e^{i{\bf q}\boldsymbol\rho}\big(1-e^{-i{\bf qs}}\big),
\label{E_pair}
\end{equation}
where $\Omega^2=\omega^2/\gamma^2+\omega_p^2$. Substituting it to (\ref{dNdw_int}) and following the procedure analogous to the one applied in \cite{TrofymenkoPRA2020} for a single particle, one obtains   
\begin{equation}
\frac{dN_k}{d\omega}=\frac{2e^2}{\pi\omega}\Bigg\{\ln\frac{2m\omega_k}{\Omega^2}-2K_0(\Omega s)+\Omega s K_1(\Omega s)-1\Bigg\}.
\label{dNdw}
\end{equation}
Here we restricted the region of integration with respect to $q$, which can be considered as the transferred momentum, by the value $q_0=\sqrt{2m\omega_k}$. It is defined as the inverse of the minimum impact parameter $\rho_0$ of the electron or positron with respect to the atomic electron. This parameter is usually chosen to equal the Bohr radius of the electron orbit at the considered atomic shell \cite{Sorensen,Williams1935} with the ionization potential $\omega_k$.

The photoionization cross section of an atom with $Z$ electrons $\sigma_{ph}=\sum\limits_k\sigma^k_{ph}$ can be expressed in terms of oscillator strength distribution function $f(\varepsilon)$ as \cite{BetheSalpeter}
\begin{equation}
\sigma_{ph}(\omega)=\frac{2\pi^2e^2Z\hbar}{mc}f(\varepsilon),
\label{sigma_f}
\end{equation} 
where, for convenience, we restored the quantities $c$ and $\hbar$. The function $f(\varepsilon)$ can be approximated by a set of $\delta$-functions corresponding to different atomic shells \cite{Talman1979} as $f(\varepsilon)=\sum\limits_k F_k\delta(\varepsilon-\varepsilon_k)$, where $\varepsilon=\hbar\omega$,  $\varepsilon_k=\hbar\omega_k$ and $F_k=Z_k/Z$. Here $Z_k$ is the number of electrons at the $k$-th shell. Substituting (\ref{dNdw}) and (\ref{sigma_f}) to (\ref{sigma_d}) and considering $\omega$ as the energy $\varepsilon$ transferred during the collision, at $\gamma\gg\omega_k/\omega_p$ for each $k$, one obtains the distant collision contribution to the differential cross section of $e^+e^-$ pair energy transfer to an atomic electron:
\begin{equation}
\begin{aligned}
\frac{d\sigma_d}{d\varepsilon}=&\frac{4\pi e^4}{m}\sum\limits_k\frac{F_k}{\varepsilon_k}\Bigg\{\ln\frac{2m\varepsilon_k}{\omega_p^2}-2K_0(\omega_p s)+\omega_p s K_1(\omega_p s)\\
&-1\Bigg\}\delta(\varepsilon-\varepsilon_k).
\end{aligned}
\label{sigma_d_explicit}
\end{equation} 

The close collision part $d\sigma_c/d\varepsilon$ is defined by the expression (\ref{w_e}) divided by $n$, since in such collisions the atomic electron is considered as free. Following \cite{Bak1987,Talman1979}, we will apply the approximation in which the atomic electron is considered as free from the very ionization edge $\varepsilon=\varepsilon_k$, while for $\varepsilon<\varepsilon_k$ the cross section $d\sigma_c/d\varepsilon$ is set equal zero. In \cite{Bak1987} it was demonstrated that such an approximation for $d\sigma_c/d\varepsilon$, together with the model for $d\sigma_d/d\varepsilon$ as a set of $\delta$-functions, nicely work for the description of experimental results on single particle straggling functions in thin silicon detectors\footnote{Though, since in this model a single effective excitation energy $\varepsilon_k$ is applied for the $k$-th atomic shell in (\ref{sigma_d_explicit}), the model might be expected to be valid in the case when the average number of collisions experienced by the particle in the target is not very small. For silicon targets this results in the appropriate target thickness of more than several microns.}. 

In the present case it is also possible to neglect the second term in the denominator of (\ref{w_e}) and the term with $J_0$ in the nominator. Concerning the term with $J_0$, this might seem inapplicable, since presently we consider the cross section at energies $\varepsilon>\varepsilon_k$, but not only $\varepsilon\gg\varepsilon_k$, as we did in the previous section. However, taking into account this term for $\varepsilon\sim\varepsilon_k$ would result in violation of self-consistency of our model since in this region of $\varepsilon$, as noted, the cross section $d\sigma_c/d\varepsilon$ is applied in the form, typical for free atomic electrons, which it has at $\varepsilon\gg\varepsilon_k$. Thus, $d\sigma_c/d\varepsilon$ reads:
\begin{equation}
\frac{d\sigma_c}{d\varepsilon}=\frac{4\pi e^4}{m}\sum\limits_k\frac{F_k}{\varepsilon^2}\theta(\varepsilon-\varepsilon_k),
\label{sigma_c_explicit}
\end{equation} 
where $\theta$ is the Heaviside step function. Expressions (\ref{sigma_d_explicit}) and (\ref{sigma_c_explicit}) lead to the following formula for the mean value of the pair ionization loss in the target:
\begin{equation}
\begin{aligned}
E_{AV}(s)&=nx\int\limits_0^{\varepsilon_\textrm{max}}\frac{d\sigma}{d\varepsilon}d\varepsilon=\eta\bigg\{\ln\frac{2m\varepsilon_\textrm{max}}{\omega_p^2}-2K_0(\omega_p s)\\
&+\omega_p s K_1(\omega_p s)-1\bigg\}.
\end{aligned}
\label{E_av_1}
\end{equation}
It differs from the result (\ref{E_av}) by the term $[\omega_p s K_1(\omega_p s)-1]$ in braces. For $s<\omega_p^{-1}$, where the Chudakov effect is the most significant, this term is close to zero. For $s\gg\omega_p^{-1}$ it tends to $-1$ and remains small compared to the logarithmic term in (\ref{E_av_1}). Thus, the result for $E_{AV}$, obtained with the use of the equivalent photon method, is very close to the result of \cite{BerestetskiiGeshkenbain} and the derived expression for $d\sigma/d\varepsilon$ can be considered as a valid approximation of the exact interaction cross section. Nevertheless, in order to be completely compatible with our consideration in the previous section, where the expression (\ref{E_av}) for $E_{AV}$ was applied, we will neglect the small term $[\omega_p s K_1(\omega_p s)-1]$ in (\ref{sigma_d_explicit}), which results in neglecting the same term in (\ref{E_av_1}), and apply the following expression for the probability of the pair energy loss $w(\varepsilon)=nd\sigma/d\omega$:  
\begin{equation}
\begin{aligned}
w(\varepsilon)=&\frac{\eta}{x}\sum\limits_k\Bigg[\frac{F_k}{\varepsilon_k}\Bigg\{\ln\frac{2m\varepsilon_k}{\omega_p^2}-2K_0(\omega_p s)\Bigg\}\delta(\varepsilon-\varepsilon_k)\\
&+\frac{F_k}{\varepsilon^2}\theta(\varepsilon-\varepsilon_k)\Bigg].
\end{aligned}
\label{w_explicit}
\end{equation} 
It differs from the corresponding expression for single particles (for $\gamma\gg\varepsilon_k/\omega_p$ for each $k$), derived in \cite{Bak1987}, by the substitution $\xi\to\eta$ and the term with $K_0$ in braces.

In order to obtain the expression for $f(x,\Delta)$ it is necessary to substitute (\ref{w_explicit}) into (\ref{f_general}) and calculate the corresponding integral. Following the procedure applied in \cite{Talman1979} for single particles, one obtains
\begin{equation}
f(x,\Delta)=\frac{1}{\pi\eta}\int\limits_0^{+\infty}dy\exp\Bigg\{-\frac{\pi y}{2}+g(y,\eta)\Bigg\}\cos\Bigg[\frac{y\Delta}{\eta}+h(y,\eta)\Bigg],
\label{f_Talman}
\end{equation}
where
\begin{equation}
\begin{aligned}
&g(y,\eta)=\sum_k\bigg\{yF_k\textrm{Si}\big(y\varepsilon_k/\eta\big)-N_k\big[1-\cos\big(y\varepsilon_k/\eta\big)\big]\bigg\},\\
&h(y,\eta)=\sum_k\bigg\{yF_k\textrm{Ci}\big(y\varepsilon_k/\eta\big)-N_k\sin\big(y\varepsilon_k/\eta\big)\bigg\},
\end{aligned}
\label{g_h_y_eta}
\end{equation}
Si and Ci are the sine and cosine integral functions, defined in such way that at $t\to+\infty$: $\textrm{Si}(t)=\pi/2$, $\textrm{Ci}(t)=0$. The function (\ref{f_Talman}) has the same structure as the one derived in \cite{Talman1979} for single particles, but differs from it by the substitution $\xi\to\eta$ and the expression for the mean number of collisions $N_k$ with the atomic electrons at $k$-th shell, which is presently
\begin{equation}
N_k=\eta\frac{F_k}{\varepsilon_k}\Bigg\{\ln\frac{2m\varepsilon_k}{\omega_p^2}-2K_0(\omega_p s)+1\Bigg\}.
\label{N_k}
\end{equation}
At large target thickness $x$, when $\eta\gg\varepsilon_k$ for each $k$, the function (\ref{f_Talman}) coincides with (\ref{f_Landau}).

\begin{figure}[h!]
	\centering
	\begin{minipage}[t]{.49\linewidth}
		\centering
		\center{\includegraphics[width=0.99\linewidth]{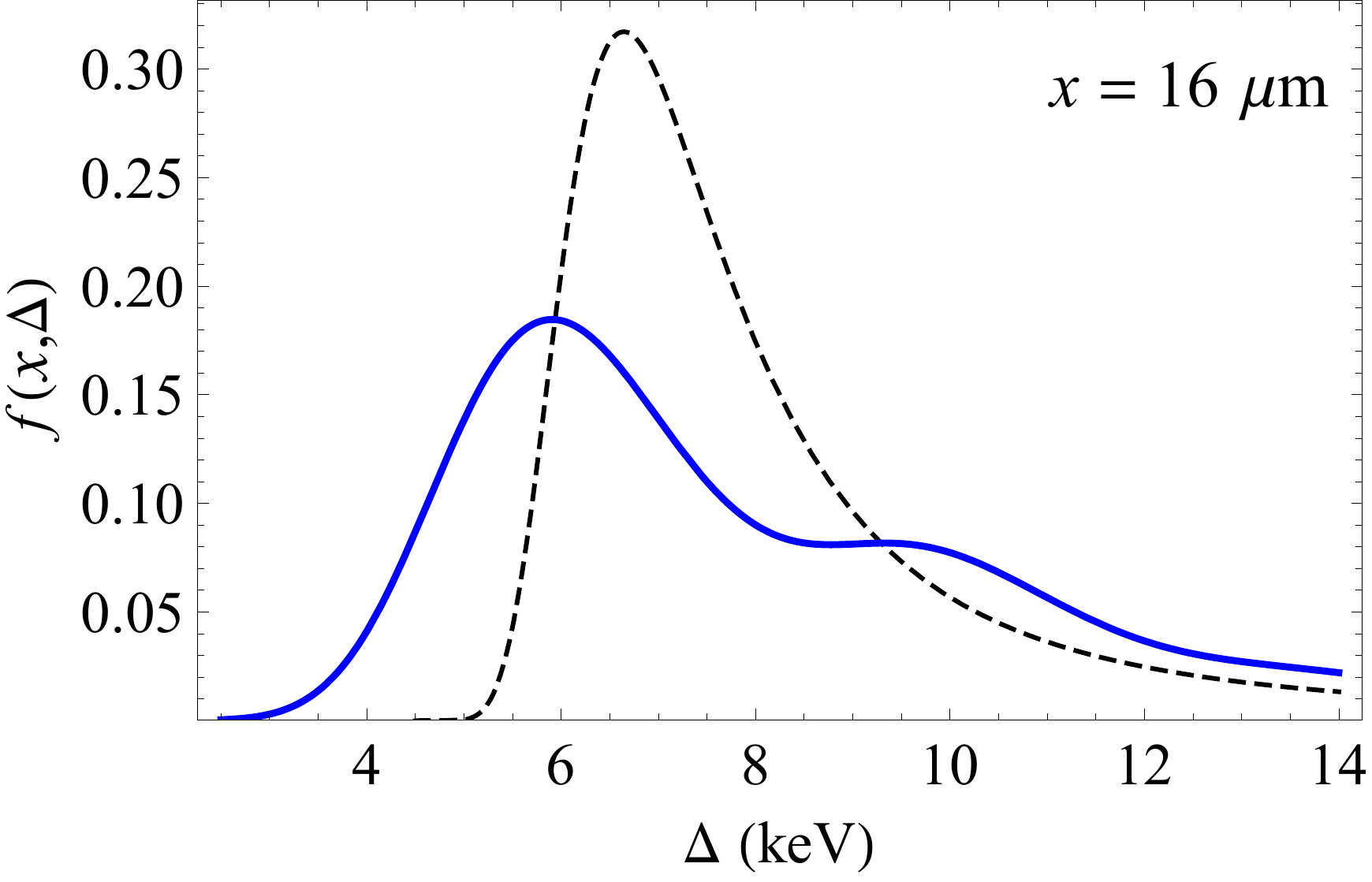}}
	\end{minipage}
	\hfill
	\begin{minipage}[t]{.49\linewidth}
		\centering
		\center{\includegraphics[width=0.99\linewidth]{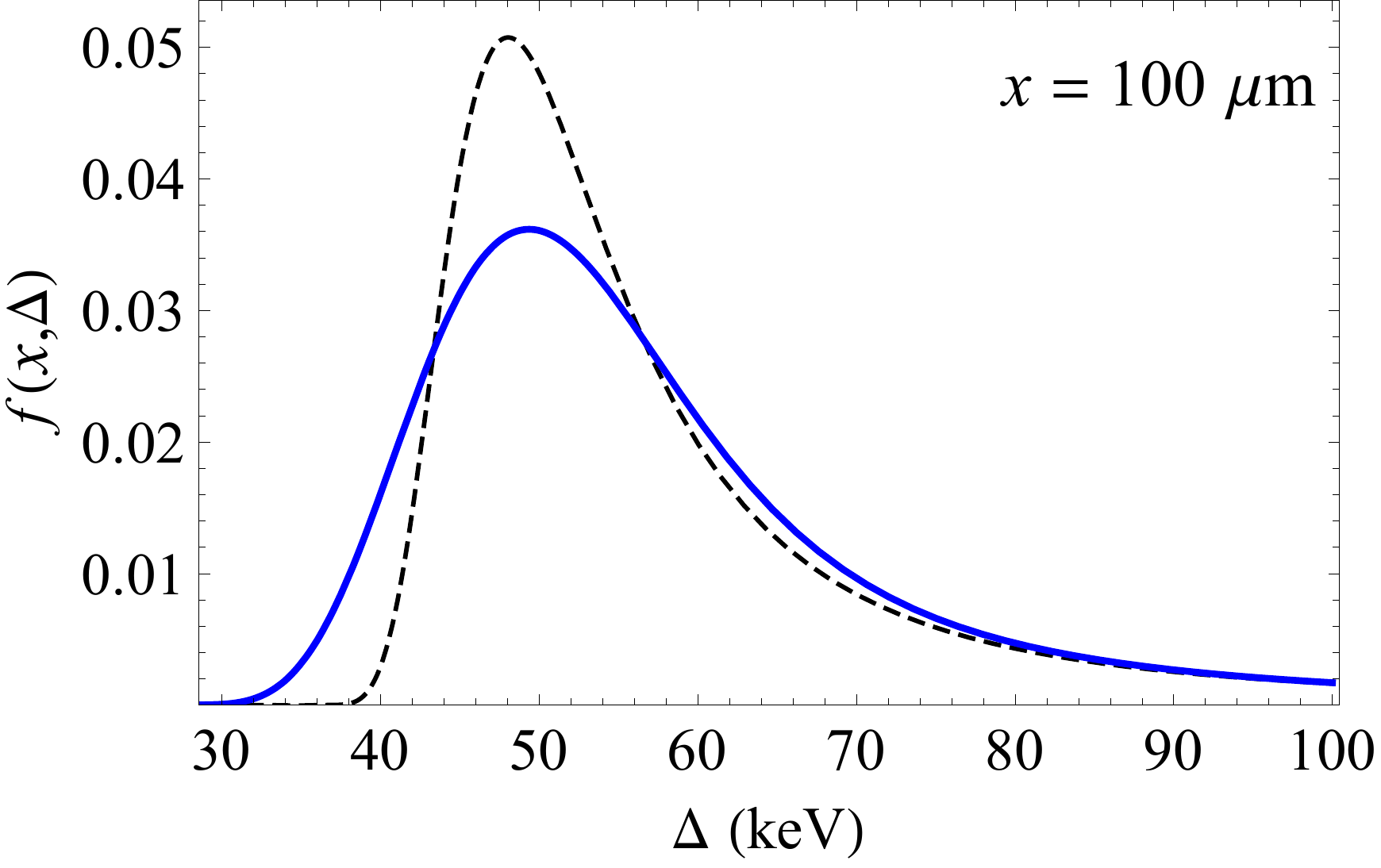}}
	\end{minipage}
	\caption{Straggling functions of a high-energy $e^+e^-$ pair for two values of the target thickness $x$. Solid lines -- calculation on the basis of (\ref{f_Talman}), dashed lines -- on the basis of (\ref{f_Landau}). Separation between the particles in the pair is $s=0.5\omega_p^{-1}$.}
	\label{fig1} 
\end{figure}

For numerical estimations we apply the following effective values of $\varepsilon_k$ for K, L and M atomic shells in silicon targets \cite{Bak1987}: $\varepsilon_K=4033$ eV, $\varepsilon_L=241$ eV and $\varepsilon_M=17$ eV. Here also $F_K=2/14$, $F_L=8/14$ and $F_M=4/14$. Fig.~\ref{fig1} shows the examples of straggling functions of $e^+e^-$ pairs in silicon targets of various thickness calculated on the basis of (\ref{f_Landau}) and (\ref{f_Talman}). The example of a target with $x=16~\mu$m is chosen since a silicon detector of this thickness has been applied in the experiment \cite{Vircus,Thomsen2011} discussed above. We see that, like in the case of single particles, in thin targets the distribution (\ref{f_Talman}) is broader than (\ref{f_Landau}). For not very small $x$ the maximum positions of these distributions do not noticeably differ, which allows applying the expression (\ref{E_mp}) in this case as well. This difference, however, becomes a bit larger with the decrease of $x$. At small $x$ a hump appears in the distribution (\ref{f_Talman}) on the right of its maximum, which is associated with the small probability of collision with K-shell electrons in this case (for details see \cite{Talman1979}). 

\begin{figure}
	\begin{center}
	\includegraphics[width = 75mm]{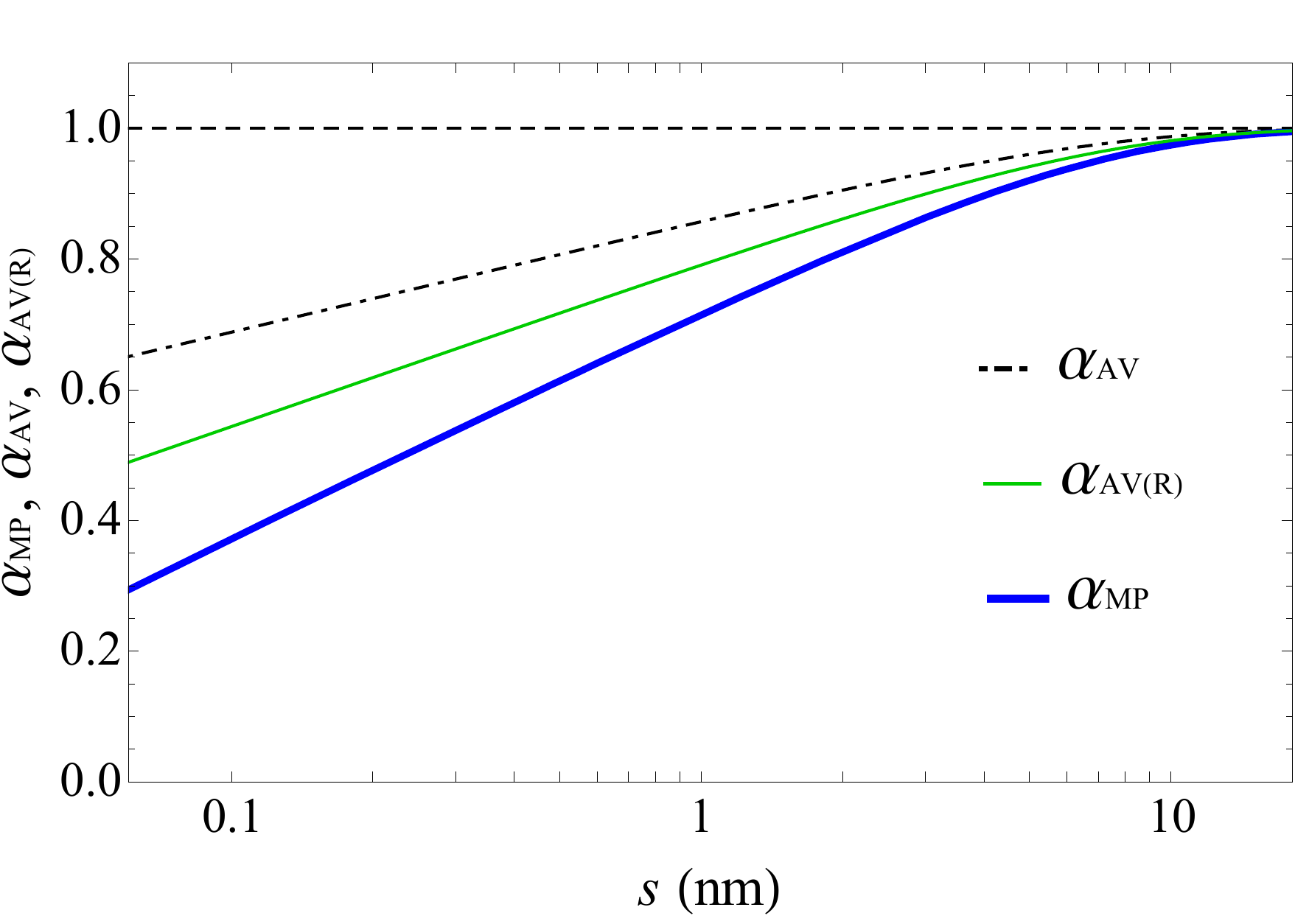}
	\caption{\label{fig2} Dependence of the relative value of $E_{MP}$ (thick solid line), $E_{AV}$ (dot-dashed line) and $E_{AV(R)}$ (thin solid line) on separation between the electron and positron in a silicon target of 16 $\mu$m thickness; $\alpha_{AV}$ is estimated for the pair energy of $\mathcal{E}=1$ GeV.}  
	\end{center}
\end{figure}

Fig.~\ref{fig2} demonstrates the dependence of ratios $\alpha_{MP}$, $\alpha_{AV}$ and $\alpha_{AV(R)}$ on $s$ for the target thickness of $16~\mu$m. The quantity $\alpha_{MP}$ is calculated numerically with the use of expression (\ref{f_Talman}). It turns out that the obtained dependence for $\alpha_{MP}$ in this case almost coincides with the one, calculated on the basis of (\ref{E_mp}) for this thickness. Thus, one concludes that although expression (\ref{E_mp}) does not provide an accurate result for the absolute value of $E_{MP}$ in sufficiently thin targets, it is still applicable here for calculation of the relative suppression of $E_{MP}$ compared to its value at $s\gg\omega_p^{-1}$. Following \cite{Vircus}, for the considered target thickness we apply the value $\varepsilon_R=100$ keV for calculation of $E_{AV(R)}$ and estimation of the quantity $\alpha_{AV(R)}$. The figure shows that the magnitude of Chudakov effect (i. e. the magnitude of the pair ionization loss suppression) is different for $E_{MP}$ and $E_{AV}$, as well as for $E_{MP}$ and $E_{AV(R)}$. This difference grows with the decrease of separation $s$ between the particles in the pair. For the smallest separations we presently consider, which are on the order of interatomic distance $a_0\sim0.1$ nm, the relative difference $(\alpha_{AV}-\alpha_{MP})/\alpha_{MP}$ reaches about 85\% for the chosen target thickness, while the value of $(\alpha_{AV(R)}-\alpha_{MP})/\alpha_{MP}$ is around 45\%. 

Fig.~\ref{fig2} shows that the value $E_{MP}$ is stronger suppressed due to Chudakov effect than $E_{AV(R)}$. As noted, in the experiment \cite{Vircus,Thomsen2011} there was registered a stronger suppression of the pair ionization loss, than it was expected from the available theories for $E_{AV(R)}$. Thus, the application of the theory, developed in the present letter, to the analysis of measurements \cite{Vircus,Thomsen2011} may improve (at least partially) the coincidence between the experimental results and the theoretical predictions. Though, as shown in \cite{Thomsen2011}, a comprehensive analysis of the discussed experiment requires taking into account such factors as angular and energy distribution of particles in the pair, their multiple scattering and uncertainty in the position of the pair creation point, which were quite noticeable under the conditions of this experiment. Such an analysis is beyond the scope of the present letter and could be performed elsewhere. For some discussion of this experiment see also \cite{TrofymenkoPLA2013}.

\section{Conclusion}
\label{Conclusions}

In this letter the straggling function of a high-energy electron-positron pair is calculated under the condition of manifestation of the Chudakov effect in the pair ionization loss. The expression for the most probable value of the pair ionization loss $E_{MP}$ is derived. The consideration is made both with the use of approach, which neglects the effects of atomic electron binding, valid in the targets of moderate thickness, and with the use of a more rigorous approach, which allows taking into account such effects and is valid in much thinner targets as well. It is shown that in sufficiently thin targets the value of $E_{MP}$ can be noticeably stronger suppressed due to Chudakov effect than the total $E_{AV}$ or restricted $E_{AV(R)}$ mean value of the pair ionization loss. The approach, developed in Sec.~\ref{sec3}, could be also applied for the study of straggling functions of larger groups of high-energy particles in very thin detectors in the case when interference effects are significant in the particle energy loss. 

\section*{Acknowledgments}
The work was partially supported by the National Academy of Sciences of Ukraine (budget program ``Support for the Development of Priority Areas of Scientific Research'' (6541230); projects 0121U111556 and 0121U111839).

\bibliographystyle{elsarticle-num-names} 
\bibliography{references}





\biboptions{sort&compress}
\end{document}